\documentclass[epj]{svjour}
\newcommand{\nc}{\newcommand}
\nc{\beq}{\begin{equation}}
\nc{\eeq}{\end{equation}}
\nc{\beqa}{\begin{eqnarray}}
\nc{\eeqa}{\end{eqnarray}}

\def\gsim{\mathrel{\rlap{\lower4pt\hbox{\hskip1pt$\sim$}}
    \raise1pt\hbox{$>$}}}       %greater than or approx. symbol
\usepackage{graphics}
\begin{document}
\title{On the Precision of a Length Measurement}
\author{Xavier Calmet \inst{1} % etc
}                     % Do not remove
\institute{Universit\'e Libre de Bruxelles, Service de Physique Th\' eorique,  CP 225, Boulevard du Triomphe, B-1050 Bruxelles, Belgique}
\date{Received: date / Revised version: date}
\abstract{
We show that quantum mechanics and general relativity imply the existence of a minimal length.  To be more precise, we show that no
operational device subject to quantum mechanics, general relativity and causality could exclude the discreteness of spacetime on lengths shorter than the Planck length. We then consider the fundamental limit coming from quantum mechanics, general relativity and causality  on the precision of the measurement of a length. 
\PACS{
      {PACS-key}{04.20.-q}   \and
      {PACS-key}{03.65.-w}
     } % end of PACS codes
} %end of abstract
\maketitle
\section{Introduction}
\label{intro}
Twentieth century Physics has been a quest for unification. The unification of quantum mechanics and special relativity required the introduction of quantum field theory. The unification of magnetism and electricity led to electrodynamics, which was unified with the weak interactions into the electroweak interactions. There are good reasons to believe that the electroweak interactions and the strong interactions originate from the same underlying gauge theory: the grand unified theory. If general relativity is to be unified with a gauge theory, one first needs to understand  how to unify general relativity  and quantum mechanics, just like it was first necessary to understand how to unify quantum mechanics and special relativity before three of the forces of nature could be unified. The aim of this paper is much more modest, we want to understand some of the features of a quantum mechanical description of general relativity using some simple tools from quantum mechanics and general relativity. In particular we shall show that if quantum mechanics and general relativity are valid theories of nature up to the Planck scale, they imply the existence of a minimal length in nature. 

We shall address two questions: Is there a minimal length in nature and is there a fundamental limit on the precision of a distance measurement? The first question will be addressed in the second section while the second will be considered in the third section. 

The usual approach to address the question of a minimal length is to do a scattering thought experiment \cite{minlength}, i.e. one studies the high energy regime of the scattering  and finds that one cannot measure a length shorter than the Planck length. Here we shall argue that this is not enough to exclude a discreteness of spacetime with a lattice spacing shorter than the Planck length. The key new idea is a precise definition of a measurement of a distance and in particular the fact that such a measurement involves actually two measurements.

We then apply our framework to the old thought experiment of Salecker and Wigner \cite{Salecker} and show that contractive states cannot beat the uncertainty due to quantum mechanics for the measurement of a length. We then conclude.

\section{Minimal Length from Quantum Mechanics and General Relativity}

In this section we review the results obtained in \cite{Calmet:2004mp}. We show that quantum mechanics and classical general relativity considered simultaneously imply the existence of a minimal length, i.e.
no operational procedure exists which can measure a
distance less than this fundamental length. The key ingredients used to
reach this conclusion are the uncertainty principle from quantum
mechanics, and gravitational collapse from classical
general relativity.

A dynamical condition for gravitational collapse is given by the hoop
conjecture \cite{hoop}: if an amount of energy
$E$ is confined at any instant to a ball of size $R$, where $R < E$,
then that region will eventually evolve into a black hole\footnote{We
use natural units where $\hbar, c$ and Newton's constant (or $l_P$)
are unity. We also neglect numerical factors of order one.}. The hoop conjecture is, as its name says, a conjecture, however it is on a firm footing. The least favorable case, i.e. as asymmetric as possible, is the one of two particles colliding head to head. It has been shown that even in that case, when the hoop conjecture is fulfilled, a black hole is formed \cite{Giddings}.

From the hoop conjecture and the uncertainty principle, we immediately deduce the
existence of a  minimum ball of size $l_P$. Consider a particle
of energy $E$ which is not already a black hole. Its size $r$ must
satisfy \beq r \gsim {\rm \bf max} \left[\, 1/E\, ,\,E \, \right]~~,
\eeq where $\lambda_C \sim 1/E$ is its Compton wavelength and $E$
arises from the hoop conjecture. Minimization with respect to $E$
results in $r$ of order unity in Planck units or $r \sim l_P$.
If the particle {\it is} a black hole, then its radius grows with mass: $r \sim E \sim 1/
\lambda_C$. This relationship suggests that an experiment designed (in
the absence of gravity) to measure a short distance $l << l_P$ will
(in the presence of gravity) only be sensitive to distances
$1/l$. 
\begin{figure}
\resizebox{0.4\textwidth}{!}{%
  \includegraphics{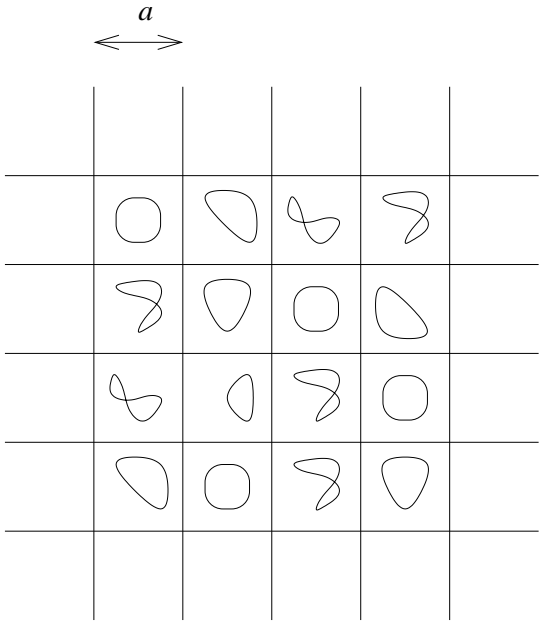}
}
%\vspace{5cm}       % Give the correct figure height in cm
\caption{We choose a spacetime lattice of spacing $a$ of the order of the Planck length or smaller. This formulation does not depend on the details of quantum gravity.}
\label{fig:1}       % Give a unique label
\end{figure}

Let us give a concrete model of minimum length. Let the position operator $\hat{x}$ have
discrete eigenvalues $\{ x_i \}$, with the separation between
eigenvalues either of order $l_P$ or
smaller.
(For regularly distributed eigenvalues with
a constant separation, this would be equivalent to a spatial lattice, see Fig. 1.)
We do not mean to imply that nature implements minimum length in this particular
fashion - most likely, the physical mechanism is more complicated, 
and may involve,
for example, spacetime foam or strings. However, our concrete 
formulation lends itself
to detailed analysis. We show below that this formulation
cannot be excluded by any gedanken experiment, which is strong evidence for the
existence of a minimum length.

Quantization of position does not by itself imply quantization of
momentum. Conversely, a continuous spectrum of momentum does not imply
a continuous spectrum of position. In a formulation of
quantum mechanics on a regular spatial lattice, with spacing $a$
and size $L$, the momentum operator has eigenvalues which are
spaced by $1/L$. In the infinite volume limit the momentum operator can have
continuous eigenvalues even if the spatial lattice spacing is kept
fixed. This means that the displacement operator \beq \label{disp}
\hat{x} (t) - \hat{x} (0) = \hat{p}(0) {t \over M} \eeq does not
necessarily have discrete eigenvalues (the right hand side of
(\ref{disp}) assumes free evolution; we use the Heisenberg picture
throughout). Since the time evolution operator is unitary the
eigenvalues of $\hat{x}(t)$ are the same as $\hat{x}(0)$. Importantly
though, the spectrum of $\hat{x}(0)$ (or $\hat{x}(t)$) is completely
unrelated to the spectrum of the $\hat{p}(0)$, even though they are
related by (\ref{disp}).  A measurement of arbitrarily small displacement
(\ref{disp}) does not exclude our model of minimum length. To
exclude it, one would have to measure a position eigenvalue $x$
{\it and} a nearby eigenvalue $x'$, with $|x - x'| << l_P$.

Many minimum length arguments are obviated by the simple observation of the minimum ball. However,
the existence of a minimum ball does not by itself preclude the
localization of a macroscopic object to very high precision.
Hence, one might attempt to measure the spectrum of $\hat{x}(0)$
through a time of flight experiment in which wavepackets of
primitive probes are bounced off of well-localised macroscopic
objects. Disregarding gravitational effects, the discrete spectrum
of $\hat{x}(0)$ is in principle obtainable this way. But,
detecting the discreteness of $\hat{x}(0)$ requires wavelengths
comparable to the eigenvalue spacing.  For eigenvalue spacing
comparable or smaller than $l_P$, gravitational effects cannot be
ignored, because the process produces minimal balls (black holes)
of size $l_P$ or larger. This suggests a direct measurement of the
position spectrum to accuracy better than $l_P$ is not possible.
The failure here is due to the use of probes with very short wavelength.

A different class of instrument, the interferometer,  is capable of measuring
distances much smaller than the size of any of its sub-components.  Nevertheless, the uncertainty principle and gravitational collapse prevent an arbitrarily accurate measurement of
eigenvalue spacing.  
First, the limit from quantum mechanics. Consider
the Heisenberg operators for position $\hat{x} (t)$ and momentum
$\hat{p} (t)$ and recall the standard inequality \beq \label{UNC}
(\Delta A)^2 (\Delta B)^2 \geq  ~-{1 \over 4} ( \langle [
\hat{A}, \hat{B} ] \rangle )^2 ~~.
\eeq Suppose that the
position of a {\it free} test mass is measured at time $t=0$
 and {\em again} at a later time.
The
position operator at a later time $t$ is \beq \label{P} \hat{x}
(t) = \hat{x} (0) ~+~ \hat{p}(0) \frac{t}{M}~~. \eeq
We assume a free particle Hamiltonian here for simplicity, but the argument can be generalized \cite{Calmet:2004mp}. The commutator between the position operators at $t=0$ and $t$
is \beq [ \hat{x} (0), \hat{x} (t)] ~=~ i {t \over M}~~,
\eeq so using (\ref{UNC}) we have \beq \vert \Delta x (0) \vert
\vert \Delta x(t) \vert \geq \frac{t}{2M}~~.\eeq
We see that at least one of the uncertainties $\Delta x(0)$ or $\Delta x(t)$
must be larger than of order $\sqrt{t/M}$.
As a measurement of the discreteness of $\hat{x}(0)$
requires {\em two} position measurements,
it is limited by the greater of $\Delta x(0)$ or $\Delta x(t)$,
that is, by $\sqrt{t/M}$, 
 \beq \label{SQL} \Delta x \equiv {\rm \bf max}\left[
 \Delta x(0), \Delta x(t) \right]
 \geq
\sqrt{ t \over 2 M }~~, \eeq where $t$ is the time over
which the measurement occurs and $M$ the mass of the object whose
position is measured. In order to push $\Delta x$ below $l_P$, we
take $M$ to be large.  In order to avoid
gravitational collapse, the size $R$ of our measuring device must
also grow such that $R > M$. However, by causality $R$ cannot
exceed $t$. Any component of the device a distance greater than
$t$ away cannot affect the measurement, hence we should not
consider it part of the device. These considerations can be
summarized in the inequalities \beq \label{CGR} t > R > M
~~.\eeq Combined with (\ref{SQL}), they require $\Delta x
> 1$ in Planck units, or \beq \label{DLP} \Delta x > l_P~. \eeq 

Notice that the considerations leading to (\ref{SQL}), (\ref{CGR})
and (\ref{DLP}) were in no way specific to an interferometer, and
hence are {\it device independent}. We repeat: no device subject
to quantum mechanics, gravity and causality can exclude the quantization
of position on distances less than the Planck length.

It is important to emphasize that we are deducing a minimum length
which is parametrically of order $l_P$, but may be larger or
smaller by a numerical factor.  This point is relevant to the
question of whether an experimenter might be able to transmit the
result of the measurement before the formation of a closed trapped
surface, which prevents the escape of any signal. If we decrease
the minimum length by a numerical factor, the inequality
(\ref{SQL}) requires $M >> R$, so we force the experimenter to
work from deep inside an apparatus which has far exceeded the
criteria for gravitational collapse (i.e., it is much denser than
a black hole of the same size $R$ as the apparatus). For such an
apparatus a horizon will already exist before the measurement
begins. The radius of the horizon, which is of order $M$, is very
large compared to $R$, so that no signal can escape.

An implication of our result is that there may only be a finite number of degrees of freedom per unit volume in our universe - no true continuum of space or time. Equivalently, there is only a finite amount of information or entropy in any finite region our universe.

One of the main problems encountered in the quantization of gravity is a proliferation of divergences coming from short distance fluctuations of the metric (or graviton). However, these divergences might only be artifacts of perturbation theory: minimum length, which is itself a non-perturbative effect, might provide a cutoff which removes the infinities. This conjecture could be verified by lattice simulations of quantum gravity (for example, in the Euclidean path integral formulation), by checking to see if they yield finite results even in the continuum limit.

\section{Limits on the Measurement of Large Distances from Fundamental Physics}

In the section, we study whether quantum mechanics and general relativity can limit the precision of a distance measurement. In order to address this question we shall reconsider the thought experiment first proposed by Salecker and Wigner almost 50 years ago. In order to measure the distance $l$ we shall consider a clock which emits a light ray at a time t=0. The clock will suffer a recoil from the emission of the light ray which induces a position uncertainty in the position of the clock $x(0)$. The mirror which is at a distance $l$ of the clock will reflect the light ray which is reabsorbed at a time $t$ at $x(t)$ by the clock, again there is  a recoil effect and the position of the clock will have some uncertainty (see Fig. 2).  
\begin{figure}
\resizebox{0.45\textwidth}{!}{%
  \includegraphics{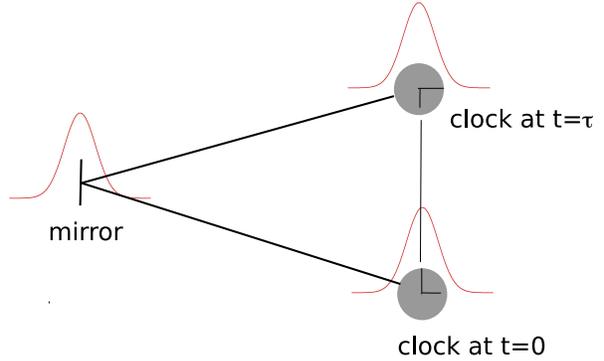}
}
%\vspace{5cm}       % Give the correct figure height in cm
\caption{Salecker and Wigner thought experiment to measure a length. A clock emits a light beam at a time $t=0$ which is reflected by a mirror and reabsorbed by the clock at a later time $t=\tau$. Quantum mechanics implies a spread of the wave function of the clock and of the mirror.}

\label{fig:2}       % Give a unique label
\end{figure}

Consider the Heisenberg operators for position $\hat{x} (t)$ and momentum
$\hat{p} (t)$ and recall the standard inequality \beq \label{UNC2}
(\Delta A)^2 (\Delta B)^2 \geq  ~-{1 \over 4} ( \langle [
\hat{A}, \hat{B} ] \rangle )^2 ~~.
\eeq Suppose that the
position of a {\it free} test mass is measured at time $t=0$
 and {\em again} at a later time.
The
position operator at a later time $t$ is \beq \label{P1} \hat{x}
(t) = \hat{x} (0) ~+~ \hat{p}(0) \frac{t}{M}~~. \eeq
The commutator between the position operators at $t=0$ and $t$
is
 \beq 
 [ \hat{x} (0), \hat{x} (t)] ~=~ i {t \over M}~~,
\eeq 
so using (\ref{UNC2}) we have 
\beq 
\vert \Delta x (0) \vert
\vert \Delta x(t) \vert \geq \frac{t}{2M}~~.
\eeq
Since the total uncertainty for the measurement of the distance $l$ is given by the sum of the uncertainties of $x(0)$ and $x(t)$ we find:
\beq 
\delta l \sim \sqrt{ \frac{t}{2M}}~~.
\eeq
Note that we are not forced to take the mass of the clock to be large like in the previous section. There are actually two options, one is to allow the mass of the clock to grow at the same rate as $t$, the time necessary for the measurement in which case we have
\beq 
\delta l \sim  1~~,
\eeq
or
\beq 
\delta l \sim  l_p~~.
\eeq
The other option is to consider a fixed, finite, mass. This case applies to e.g. the measurement of a distance performed with an interferometer such as LIGO \cite{LIGO}. The mass is at most the mass of the region of spacetime which feels one wavelength of the gravitational wave. In that case, the standard quantum limit \cite{SQL}  applies, and this is the well-known statement that LIGO operates at the standard quantum limit. Note that contractive states \cite{Yuen,Ozawa} cannot help to beat the standard quantum limit in a parametric manner. Again as in \cite{Calmet:2004mp} the reason is that we need two measurements. Contractive states allow to make $\delta x(t)$ very small at the price of losing all the information about the uncertainty of $x(0)$ (see appendix).

\section{Conclusions}

In this work we have shown that quantum mechanics and classical general relativity considered simultaneously imply the existence of a minimal length, i.e.
no operational procedure exists which can measure a
distance less than this fundamental length. The key ingredients used to
reach this conclusion are the uncertainty principle from quantum
mechanics, and gravitational collapse from classical
general relativity. Furthermore we have shown that contractive states cannot be used to beat the limit obtained by Salecker and Wigner  on the precision of a measurement of a length. Note that in that case we are not forced to consider very massive objects and thus the gravitational collapse condition does not necessarily provide a bound. If we are forced to consider very massive objects, then the best precision for the measurement of a length which can be archived is the minimal length itself. Our results have deep consequences for the detectability of quantum foam using astrophysical sources \cite{Ng:1999se,Ng:1999hm,Amelino-Camelia:2000ev,Lieu:2002jt,Aloisio:2002ed,Ng:2003ag,Lieu:2003ee,Christiansen:2005yg}. This however goes beyond the scope of this paper and shall be considered elsewhere.

\bigskip
%%%%%%%%%%%%%%%%%%%%%%%%%%%%%%%%%%%%%%%%%%%%%%%%%%%%%%%%%%%%%%%%%
%%%
%%%                   ACKNOWLEDGMENTS
%%%
%%%%%%%%%%%%%%%%%%%%%%%%%%%%%%%%%%%%%%%%%%%%%%%%%%%%%%%%%%%%%%%%%
\section*{Acknowledgments}
I would like to thank G. 't Hooft and A. Zichichi for the wonderful Erice Summer School on Subnuclear Physics they have organized and where part of this work was done. It is with great pleasure that I acknowledge that the results of the third section were worked out during this summer school with S. Hsu. I am very grateful to A. Zichichi for the financial support that made my participation to this school possible as well as to H. Fritzsch who had nominated me for this school.
Finally, I would like to thank F.R. Klinkhamer for enlightening discussions and for drawing my attention to the work of Salecker and Wigner.  I would like to thank M. Ozawa for a helpful communication and for sending me a copy of his work on contractive states. This work was supported in part by the IISN and the Belgian science
policy office (IAP V/27).

\section*{Appendix: Contractive States}
Here we briefly review contractive states, following  Ozawa's original work \cite{Ozawa}.  One introduces the operator $\hat a$ defined by
\beq
\hat a = \sqrt{\frac{m \omega}{2 \hbar}} \hat x +\sqrt{\frac{1}{2 \hbar m \omega}}  i \hat p.
\eeq
The quantization of $\hat x$ and $\hat p$ implies $[\hat a, \hat a ^\dagger]=1$. The parameter $\omega$ is free. The twisted coherent state $|\mu\nu\alpha\omega\rangle$ is the eigenstate of $\mu \hat a + \nu \hat a^\dagger$  with eigenvalue $\mu \hat \alpha + \nu \hat \alpha^\star$. The normalization of the wave function implies $|\mu|^2 -|\nu|^2=1$. The free Hamiltonian is given by 
\begin{eqnarray}
\hat H = \hat p/2m = \hbar \omega/2 (\hat a^\dagger \hat a +\frac{1}{2} - \frac{1}{2} \hat a^2 - \frac{1}{2} \hat a^{\dagger 2} )
\end{eqnarray}
and the wave function of this state is given by
\begin{eqnarray}
\langle  x|\mu\nu\alpha\omega\rangle &=&\left (\frac{m\omega}{\pi \hbar |\mu-\nu|^2} \right )^{1/4} \times  \\ \nonumber &&
\mbox{exp}\left( - \frac{m \omega}{\pi \hbar} \frac{1+2 \xi i}{|\mu-\nu|^2 } (x-x_0)^2 + i p_0 (x-x_0)\right)
\end{eqnarray}
with $\xi=\mbox{Im}(\mu^\star\nu)$, $\alpha= (m \omega/2\hbar)^{1/2} x_0 + 1/(2\hbar m \omega)^{1/2} i p_0$ where $x_0$ and $p_0$ are real. 
The position fluctuation for a free-mass is given by:
\begin{eqnarray}
\Delta x(t)^2 = \frac{1}{4\xi} \frac{\hbar \tau}{m} + \frac{2 \hbar}{m\omega} \left (|\mu+\nu| \frac{\omega}{2} \right)^2 (t-\tau)^2
\end{eqnarray}
with
\begin{eqnarray}
\tau = \xi \hbar m/\Delta p(0)^2.
\end{eqnarray}
When $\xi>0$ the $x$-depdendent phase leads to a narrowing of $\Delta x(t)$ compared to  $\Delta x(0)$. States with the property  $\xi>0$  are called contractive states. The absolute minimum is achieved for a time $\tau$ given by
\begin{eqnarray}
\tau= \frac{2 \xi}{\omega |\mu+\nu|^2)}=\frac{ \xi \hbar m}{\Delta p(0)^2}
\end{eqnarray}
and one obtains
\begin{eqnarray}
\Delta x(\tau) = \frac{\hbar}{2 \Delta p(0)}= \frac{\Delta x(0)}{\sqrt{1+4 \xi^2}}.
\end{eqnarray}
The price to pay to make $\Delta x(\tau)$ very small, i.e. smaller than e.g. the Planck length, is to pick $\xi$ very large, which implies that $\tau$ is very large and thus $\Delta x(0)^2$ is very large as well. Keeping in mind that the measurement of a distance implies two measurements we see that it is not possible to parametrically make the uncertainty on the measurement of a distance arbitrarily small. This is equivalent to the statement that LIGO and similar interferometers operate at the quantum limit, one can beat by some small factor the standard quantum limit, but it cannot be beaten parametrically.

\end{document}